\documentclass[aps, prl, 10pt, twocolumn, superscriptaddress,noshowpacs, preprintnumbers, longbibliography, bibnotes,hyperref,floatfix]{revtex4-1}

\usepackage[colorlinks=true,breaklinks=true]{hyperref}
\hypersetup{allcolors=[rgb]{0.0 0.0 1.0},linkcolor=[rgb]{0.75 0.05 0.05}}
\usepackage[dvipsnames]{xcolor}
\usepackage{mathtools}
\usepackage{amsmath}
\usepackage{amsfonts}
\usepackage{amssymb}
\usepackage{bbold}
\usepackage{multirow}
\usepackage{bm}
\usepackage{hyperref}
\long\def\exclude#1{}
\usepackage{orcidlink}

\newcommand{\br}{{\bf r}}
\newcommand{\bk}{{\bf k}}

\newcommand{\vP}{{\vec{P}}}

\newcommand{\bv}{{\bf v}}

\newcommand{\tpsi}{{\tilde{\psi}}}

\newcommand{\bE}{{\bf E}}

\begin{document}

\title{Fast flavor conversions at the edge of instability in a two-beam model}

\author{Damiano F.\ G.\ Fiorillo \orcidlink{0000-0003-4927-9850}} 
\affiliation{Niels Bohr International Academy, Niels Bohr Institute,
University of Copenhagen, 2100 Copenhagen, Denmark}

\author{Georg G.\ Raffelt
\orcidlink{0000-0002-0199-9560}}
\affiliation{Max-Planck-Institut f\"ur Physik (Werner-Heisenberg-Institut),  Boltzmannstr.~8, 85748 Garching, Germany}

\begin{abstract}
A dense neutrino gas exhibiting angular crossings in the electron lepton number is unstable and develops fast flavor conversions. Instead of assuming an unstable configuration from the onset, we imagine that the system is externally driven toward instability. We use the simplest model of two neutrino beams initially of different flavor that either suddenly appear or one or both slowly build up. Flavor conversions commence well before the putative unstable state is fully attained, and the final outcome depends on how the system is driven. The system generally sticks to the closest state that is linearly stable, a conclusion that we prove for the first time using quasi-linear theory. Our results suggest that in an astrophysical setting, one should focus less on flavor instabilities in the neutrino radiation field and more on the external dynamics that leads to the formation of the unstable state.
\end{abstract}

\date{\today}

\maketitle

{\bf\textit{Introduction.}}---In dense neutrino environments, refractive flavor exchange can occur on timescales much faster than vacuum oscillations. These \textit{fast flavor conversions (FFCs)} \cite{Sawyer:2015dsa, Chakraborty:2016lct, Izaguirre:2016gsx, Tamborra:2020cul, Fiorillo:2024wej, Richers:2022zug, Patwardhan:2022mxg} have received widespread attention for their potential impact on the evolution of sources such as core-collapse supernovae (CCSNe) and neutron-star mergers (NSMs) \cite{Ehring:2023lcd, Ehring:2023abs, Nagakura:2023mhr, Wu:2017drk, Volpe:2023met}. A concrete evaluation of their role is challenging, because FFCs happen on very short length and time scales. One strategy simply relies on consistency; one starts with a putative state obtained in the simulation \textit{without} FFCs and investigates its subsequent behavior. If the system is stable, it was justified to ignore FFCs. 

However, the very interest in this subject derives from the opposite finding, i.e., FFC instabilities seem to be generic in numerical CCSN and NSM simulations \hbox{\cite{Abbar:2018shq,Li:2021vqj,Abbar:2019zoq,Nagakura:2019sig,Abbar:2020qpi,Nagakura:2021hyb,Wu:2017qpc}}. Many recent studies try to understand what would be the self-consistent solution in that they follow the post-instability evolution of a simplified system of ``neutrinos in a box,'' i.e., the temporal evolution of a homogeneous neutrino gas with an FFC instability as initial condition~\cite{Martin:2021xyl,Zaizen:2023ihz,Zaizen:2022cik,Xiong:2023vcm,Grohs:2022fyq,Richers:2021xtf,Richers:2022bkd,Bhattacharyya:2020jpj,Bhattacharyya:2022eed,Wu:2021uvt,Martin:2019gxb,Abbar:2021lmm,Duan:2021woc,Cornelius:2023eop,Fiorillo:2024fnl}. However, the realism of such exercises is unclear in the sense that the unstable state would not form in the first place, as also discussed in Ref.~\cite{Johns:2024dbe}. A more realistic formulation of the problem might be to start with a stable configuration and \textit{slowly} drive it towards instability by some external agent. Presumably, as soon as a weak instability first appears, FFCs will prevent its full development. This formulation begs the question: does the evolution toward the putative instability affect the final outcome?

We here perform the first study to address this question in the form of the simplest conceivable toy model. It consists of two neutrino beams moving in opposite directions with different flavor content, specifically one of them initially consisting of $\nu_e$ and the other of $\nu_\mu$. The final outcome is expected to be some space-time varying mixture, representing some sort of flavor equipartition under the constraints of conserved quantities, at least in the sense of an ensemble average for different initial seeds for the unstable modes. The conventional approach to the problem amounts to the \textit{sudden} appearance of the unstable configuration. We compare this with the more realistic setting where one or both of the beams are built by external feeding over timescales much slower than the instability growth rate. 

Our results show that in this second scenario the evolution is relatively simple to understand within the framework of quasi-linear theory of the instability~\cite{Vedenov1962quasi, drummond1961non, Drummond:1964}, in which the small transverse oscillations of the polarization vectors are treated linearly, but their non-linear feedback on the space-averaged configuration is accounted for. If slowly driven, the system on average tends to stick to the closest stable configuration along its entire evolution. In turn, this implies that the final outcome can systematically differ from the conventional sudden approach, and may depend on the driving mechanism of the instability.

{\bf\textit{Two-beam model.}}---A simple system allowing for fast flavor exchange consists of two counter-moving beams, which we describe as a right-moving (R) beam with velocity $v=+1$ and a left-moving (L) one with $v=-1$. The particle and flavor content is represented by the density matrices $\rho_i$ ($i={\rm L}$ or R), normalized to a fiducial neutrino density $n_0$ such that the individual number densities are $n_i=n_0\,{\rm Tr}\,\rho_i$. These are taken to be homogeneous and therefore conserved when neutrinos stream freely. To be specific, we always use $n_0=n_{\rm R}$ and $n_{\rm L}\leq n_{\rm R}$. Moreover, we express the two-flavor content in terms of the usual polarization vectors as \smash{$\rho_i=\frac{1}{2}\bigl(P_i^0\,\mathbb{1}+\vP_i\cdot{\vec\sigma}\bigr)$}. We also use the traditional angular moments $\vP_n=\sum_v v^n \vP_v$ that here reduce to the polarization vector for the density, \smash{$\vP_0=\vP_{\rm R}+\vP_{\rm L}$}, and one for flux \smash{$\vP_1=\vP_{\rm R}-\vP_{\rm L}$}. The equations of motion (EOMs) are
\begin{subequations}\label{eq:EOM}
    \begin{eqnarray}\label{eq:EOM-R}
    \partial_t \vP_{\rm R}+\partial_r \vP_{\rm R}&=&\bigl(\vP_0-\vP_1\bigr)\times \vP_{\rm R},
    \\ \label{eq:EOM-L}
    \partial_t \vP_{\rm L}-\partial_r \vP_{\rm L}&=&\bigl(\vP_0+\vP_1\bigr)\times \vP_{\rm L},
    \end{eqnarray}
\end{subequations}
where all variables depend on $t$ and $r$. (Notice that we denote the spatial variable along the beam with $r$, whereas $x$, $y$, and $z$ denote directions in flavor space.) The neutrino-neutrino interaction strength was absorbed in the units of space and time, leaving the equations without dimensionful scale. In terms of global properties, the space-averaged density $\langle\vP_0\rangle$ is conserved, whereas the space-averaged flux $\langle\vP_1\rangle$ evolves nontrivially.

This setup is identical to the one for our recent study of energy nonconservation \cite{Fiorillo:2024fnl} and once more, we consider FFC between two flavors $\nu_e$ and $\nu_\mu$, completely avoiding antineutrinos in the discussion. As initial condition we use $P_{\rm R}^z=\zeta_{\rm R}=+1$ (meaning $\nu_e$) and $P_{\rm L}^z=\zeta_{\rm L}<0$ (meaning $\nu_\mu$) and vanishing $x$--$y$ components except for small initial seeds. The homogeneous mode of a two-beam setup is always stable, and even for multiple beams, a purely homogeneous mode evolves in a simple periodic way due to many nontrivial conserved quantities~\cite{Johns:2019izj, Fiorillo:2023mze, Fiorillo:2023hlk}. Thus, the instability necessarily entails a spontaneous breaking of homogeneity. 

\begin{figure*}
    \centering
    \includegraphics[width=0.48\textwidth]{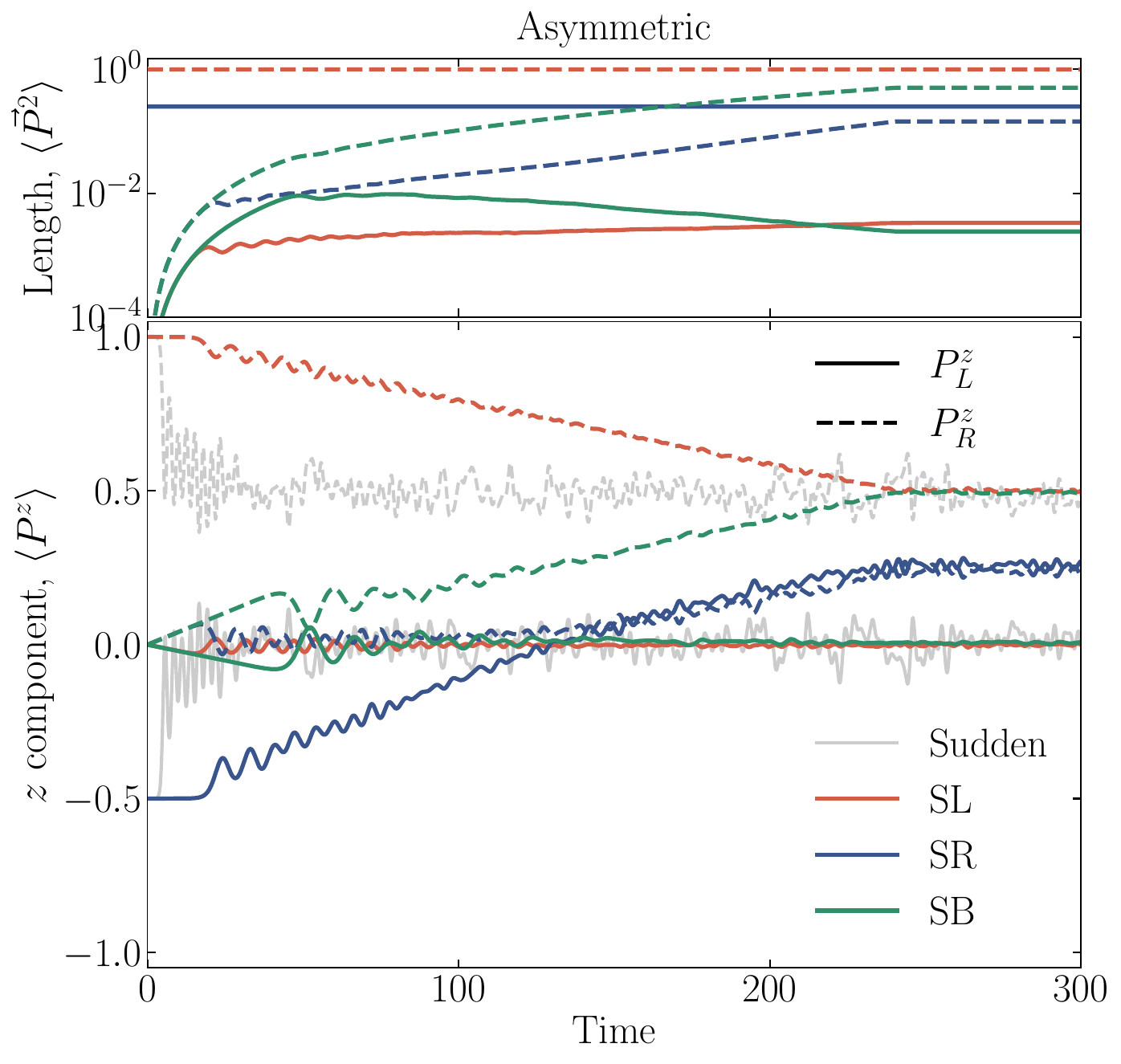}
    \includegraphics[width=0.48\textwidth]{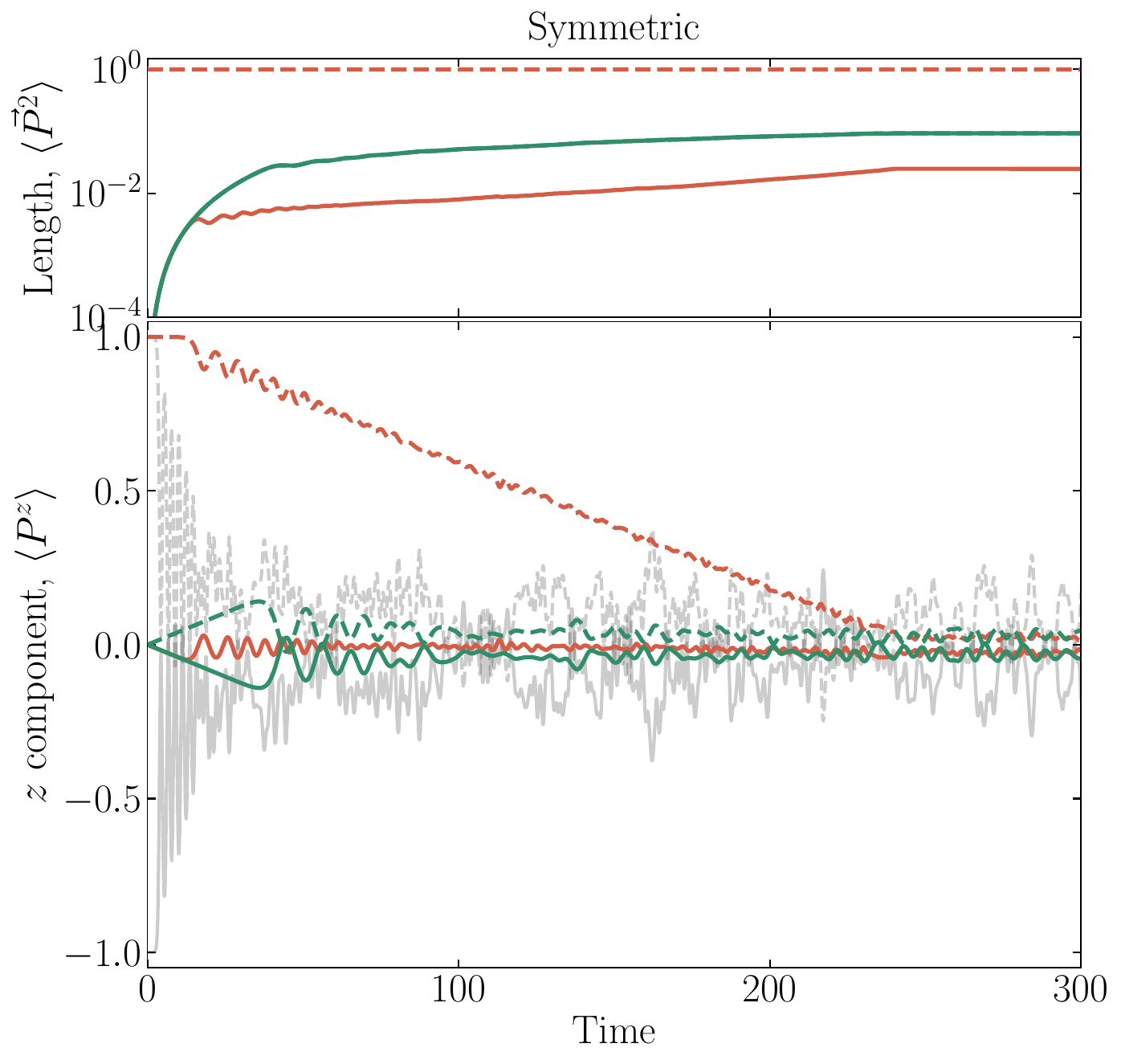}
    \caption{Evolution of box-averaged quantities, for the asymmetric (left panels) and symmetric (right panels) beam configuration, dashed lines for the R beam, solid ones for L. The sudden case in gray, the slow feeding cases SL, SR and SB in the indicated colors. The main panels show the flavor evolution in the form of the box-averaged $\langle P^z_i\rangle$, the top panels the average squared length $\langle \vP_i^2\rangle$, not shown for the sudden case where it is conserved. The blue lines in the asymmetric case show that a different asymptotic result obtains in the SR case relative to SL and SB.}
    \label{fig:main_figure}
\end{figure*}

{\bf\textit{Numerical solution for the sudden case.}}---The traditional approach was to assume an unstable configuration as an initial condition, i.e., the beams appear suddenly and then evolve. As a first example we consider this {\em sudden case} for a symmetric initial setup with $\zeta_{\rm R}=1$ and $\zeta_{\rm L}=-1$. We solve the EOMs numerically in a periodic box of length $L=300$, implying Fourier modes $k_n=2\pi n/L$, and we provide small Gaussian seeds for the modes with $n\leq100$ \cite{Fiorillo:2024fnl}. The nearly homogeneous initial system becomes inhomogeneous with disturbances cascading to ever smaller scales. However, the amplitude appearing at large $k$ is exponentially suppressed, allowing one to use a numerical scheme that damps small-scale perturbations and thus reaching larger overall integration times. We use a pseudo-spectral approach where the advection term in Eq.~\eqref{eq:EOM} is solved in Fourier space, the nonlinear term in coordinate space with a resolution of 2500 grid points. At each step we suppress all modes with $|k|>k_\mathrm{max}=17$ to avoid numerical small-scale instabilities to develop. We have tested that different values of the cutoff, even with $k_\mathrm{max}=40$, do not change the large-scale behavior of the solution. 

As expected, the two beams quickly exchange flavor as measured by the box-averaged $\langle P_{\rm L}^z\rangle$ and $\langle P_{\rm R}^z\rangle$ shown as dashed and solid gray lines in the right panel of Fig.~\ref{fig:main_figure}. Because $\langle P_{\rm L}^z\rangle+\langle P_{\rm R}^z\rangle$ is conserved, these curves are identical up to a sign. The curves keep oscillating without apparent further damping, i.e., their oscillatory power does not dissipate to ever smaller scales as mentioned earlier. Moreover, the asymptotic behavior does not approach complete equipartition---each beam retains a small excess of its original flavor.

Our second generic case is that of asymmetric beams with $\zeta_{\rm R}=1$ and $\zeta_{\rm L}=-1/2$. The solution for the sudden case is shown as dashed and solid gray lines in the left panel of Fig.~\ref{fig:main_figure}. Once more, flavor exchange is fast and at late times, the solutions oscillate around $\langle P_{\rm R}^z\rangle\simeq 1/2$ and $\langle P_{\rm L}^z\rangle\simeq 0$, with no apparent offset from these expected equilibration values. The general lesson is that the weaker beam reaches equipartition, which mirrors the conclusions obtained for continuous angular distributions in periodic boxes, namely that the nonlinear evolution entails a removal of the crossing~\cite{Zaizen:2022cik, Zaizen:2023ihz, Xiong:2023vcm, Cornelius:2023eop,Nagakura:2023jfi}. The use of periodic conditions may directly impact this conclusion~\cite{Zaizen:2023wht,Cornelius:2023eop}, although in our case the times over which the distributions relax are much shorter than the light-crossing time of the box. Thus, the periodicity is not affecting our results, while of course the choice of a quasi-homogeneous initial condition is relevant in determining the final outcome.

{\bf\textit{Slow appearance.}}---Our main interest concerns the possible changes if one or both of the beams appear slowly. To this end we assume, e.g., that initially $\vP_{\rm L}(0,r)=0$ and we include on the right-hand side of Eq.~\eqref{eq:EOM-L} a homogeneous source term $\dot P^z_{\rm L}=\zeta_{\rm L}/t_{\rm max}$ for the period $0\leq t\leq t_{\rm max}$ and zero for larger $t$ when the beam would have reached the equivalent sudden value; specifically we use $t_{\rm max}=240$. With this feeding procedure, the spatial average $\langle{P}^z_0\rangle$ is unaffected by flavor exchange and simply grows with the rate $\zeta_{\rm L}/t_{\rm max}$. We consider three different cases: (SL)~the L beam grows slowly, the R beam is present from the start. (SR)~The opposite case. (SB)~Both beams grow slowly over the same time scale $t_{\rm max}$ to their final densities $\zeta_{\rm L, R}$.

The evolution is quite different when the instability is slowly driven. Generally, as soon as the system leaves the stability region even slightly, the polarization vectors quickly saturate to the closest configuration that is linearly stable while conserving $\langle{P}^z_0\rangle$. For asymmetric beams (left panels of Fig.~\ref{fig:main_figure}) this is most easily seen for the SL case (red lines), where the L beam is slowly fed muon neutrinos, but $P^z_L$ never actually grows, oscillating around $0$ with small oscillation amplitudes (solid red line). The average properties of the final state after the feeding has stopped are similar to the sudden case, except that the asymptotic oscillations around the mean are of much smaller amplitude.

For SB, the situation is quite similar. Even though both beams are fed simultaneously, R is always stronger than L because of faster growth, and therefore, L always sticks on average to equipartition (green lines). The final configuration is similar to the SL case.

The most intriguing case is SR, where initially the R beam is empty, so it is the one that is brought on average to equipartition. When the feeding reaches a point where R would have become stronger than L, were it not for the fast conversions, the system sticks to the closest stable configuration where $\langle{P}^z_{\rm L}\rangle\simeq\langle{P}^z_{\rm R}\rangle$ and grow together. After feeding stops, this leads to a completely different final configuration with $\langle{P}^z_{\rm L}\rangle\simeq\langle{P}^z_{\rm R}\rangle\simeq0.25$. We reach the intriguing conclusion {\em that the final outcome depends not only on the putative unstable configuration, but also on the history with which that configuration is realized}. This is our main conclusion.

For symmetric beams, the SL and SR cases are analogous and in all cases, equipartition is reached except for a small offset in analogy to the sudden case.

The EOMs of Eq.~\eqref{eq:EOM} conserve $|\vP_i(r,t)|$ along trajectories, implying that for each beam, the spatial average \smash{$\langle\vP_i^2\rangle$} is conserved. In contrast, when one of the two beams is slowly driven, $\langle\vP_i^2\rangle$ never grows to the value it would have reached if FFC would not happen during injection. This behavior highlights that the density matrix for each beam represents a large ensemble of individual neutrinos with different energies and slightly different directions that decohere relative to each other while FFCs happen. We show the evolution of $\langle\vP_i^2\rangle$ in the upper panels of Fig.~\ref{fig:main_figure} and conclude that for the slowly growing beams, the average length remains very small, explaining the smaller amplitude of the final-state oscillations. In the sudden case, FFC only consists of the polarization vectors attaining random directions, whereas their length remains conserved on a trajectory.

{\bf\textit{Quasi-linear analysis.}}---Our numerical examples suggest that the system always sticks to the closest linearly stable configuration, except in the special case of symmetric beams where $\langle{P}^z_0\rangle=0$ and small deviations from this rule are seen in that the final state shows a small offset from flavor equilibration. Here we show that this simple conclusion actually descends directly from a quasi-linear treatment of the instability~\cite{Vedenov1962quasi, drummond1961non, Drummond:1964,schKT}; see our Supplemental Material~\cite{supplementalmaterial} for an introductory exposition of this framework in plasma physics.

Since in all our examples the feeding is much slower than FFC, and the numerical results suggest that it always stays close to a stable configuration, we consider our system initially in a slightly unstable configuration with no feeding. It will be convenient to frame the discussion in terms of the moments $\vP_0$ and $\vP_1$; initially, $P^z_{0,1}(t=0,z)=\zeta_{0,1}=\zeta_{\rm R}\pm\zeta_{\rm L}$. 
If $|\zeta_1|\gtrsim|\zeta_0|$, the system exhibits some unstable eigenmodes. A $k$ mode is unstable if $k_1<k<k_2$, where $k_{1,2}=\zeta_{\rm L}-\zeta_{\rm R}\mp2\sqrt{-\zeta_{\rm L}\zeta_{\rm R}}$. The corresponding eigenfrequencies are $\omega_k=\omega_0\pm i\gamma_k$, where the precession frequency $\omega_0=-\zeta_{\rm L}-\zeta_{\rm R}$ does not depend on $k$ and the growth rate is $\gamma_k=\sqrt{(k-k_1)(k_2-k)}$. In particular, the maximum growth rate is attained for $\overline{k}=\zeta_{\rm L}-\zeta_{\rm R}=(k_1+k_2)/2$ and is $\overline{\gamma}=2\sqrt{-\zeta_{\rm L} \zeta_{\rm R}}=(k_2-k_1)/2$ \cite{Fiorillo:2024fnl}. This frequency, complete with its real part and expressed in terms of $\zeta_0$ and $\zeta_1$, reads $\overline{\omega}=\omega_0+i\overline{\gamma}=-\zeta_0+i\sqrt{\zeta_1^2-\zeta_0^2}$ as mentioned earlier. Thus if $\sqrt{\zeta_1^2-\zeta_0^2}\ll |\zeta_0|$, we can consider the growth very slow and the system nearly stable. Under this condition, we may study the evolution by separating it into a background which changes slowly, over times of order $\overline{\gamma}^{-1}$, and a fluctuating part changing rapidly, over times of order $\omega_0^{-1}$. The tenet of the quasi-linear approximation we propose here is that the fluctuating part remains much smaller than the background.


In practice, we perform the separation by defining the background as the spatial average $P_{0,1}(t)=\langle P^z_{0,1}(t,r)\rangle$. In principle, one could rather define it as an ensemble average over initial conditions; if the latter are randomly sampled as in our numerical examples, such an average will be homogeneous, leading to identical results. Furthermore, the symmetry of the initial conditions ensures that the background solution has all polarization vectors aligned with the $z$ axis. In linear theory, $P_{0,1}(t)$ would be assumed to be constant; here we account for their slow change induced by the fluctuating part.

The latter is described by the transverse components $\psi_{0,1}(t,r)=P^x_{0,1}+iP^y_{0,1}$; small fluctuations could also affect the $z$ component, but they appear at higher order and are therefore neglected in quasi-linear theory. It is more convenient to use the Fourier transform $\tilde{\psi}_{0,1}(t,k)=\int_0^L dr\, \psi_{0,1}(t,r) e^{-ikr}$. The fluctuating part evolves with the linear EOM
\begin{subequations}\label{eq:fluct}
\begin{eqnarray}
    \partial_t\tpsi_0+ik\tpsi_1&=&0,\\ 
    \partial_t \tpsi_1+ik\tpsi_0&=&2i\left[P_0 \tpsi_1-P_1 \tpsi_0\right].
\end{eqnarray}    
\end{subequations}
Differently from conventional linear theory, here $P_{0,1}(t)$ evolve according to the quasi-linear equation
\begin{subequations}
\begin{eqnarray}
    \kern-2em\partial_t P_0&=&0,
    \\ 
    \kern-2em\partial_t P_1&=&i\sum_k\left[\tpsi_0(t,k) \tpsi_1^*(t,k)-\tpsi^*_0(t,k) \tpsi_1(t,k)\right].
\end{eqnarray}    
\end{subequations}
A more detailed derivation of these equations is given in the Supplemental Material~\cite{supplementalmaterial}. This approximation is in a sense reminiscent of the slow-dynamics equation we derived elsewhere~\cite{Fiorillo:2023ajs} (see also Refs.~\cite{Froustey:2020mcq,Froustey:2021azz}), where the fast dynamics admits some solution; in that case the slow dynamics was driven by collisions, whereas here it is driven by the growth of the instability.

The exact solution for $\tpsi_{0,1}$ is nontrivial, because the background evolves. However, since the growth rate is very slow, we can use the WKB approximation that each eigenmode of linear stability analysis evolve in time with the factor $e^{-i\int_0^t \omega_k(t') dt'}$, where $\omega_k(t)=\omega_0(t)+i\gamma_k(t)$ is the frequency of the eigenmode. We can keep only the unstable eigenmodes with $\gamma_k>0$, which are the only ones that grow to a considerable level. From Eq.~\eqref{eq:fluct}, for these eigenmodes we have $\tpsi_1=\omega_k \tpsi_0/k$. Substituting in the equation for $P_1$, we find
\begin{equation}
    \partial_t P_1=2\sum_k \frac{\gamma_k\mathcal{P}_k}{k},
\end{equation}
where $\mathcal{P}_k(t)=|\tpsi_0(t,k)|^2$ measures the power in the unstable eigenmode. Its evolution is then determined by the WKB approximation as
\begin{equation}
    \partial_t \mathcal{P}_k=2\gamma_k \mathcal{P}_k.
\end{equation}
These EOM completely determine the quasi-linear evolution. However, to get a qualitative feeling about their predictions, we can simplify them even further assuming that only the most unstable eigenstate contributes. As mentioned earlier and following Ref.~\cite{Fiorillo:2024fnl}, this corresponds to $\overline{k}=-P_1$ with $\gamma_{\overline{k}}=2\sqrt{P_1^2-P_0^2}$. Denoting by $\mathcal{P}=\mathcal{P}_{\overline{k}}$, we find the two coupled equations
\begin{equation}
    \partial_t P_1=-\frac{4\sqrt{P_1^2-P_0^2}\,\mathcal{P}}{P_1},
    \quad
    \partial_t \mathcal{P}=4\sqrt{P_1^2-P_0^2}\,\mathcal{P}.
\end{equation}
While this approximation has no pretense of a detailed description, given the neglect of many unstable states, it clearly shows the qualitative trend: the only fixed point is $|P_1|=|P_0|$. Thus, quasi-linear theory predicts that after initialization, the system evolves towards the closest state that is linearly stable, exactly as in the numerical experiments. 

The predicted quasi-linear evolution would be quantitatively accurate in those cases where the amplitude of the fluctuations $\psi_{0,1}$ remains small, i.e., when the instability is very weak and a small change is enough to remove it. However, even when the amplitude does grow large, quasi-linear theory correctly predicts the qualitative trend, a feature well known from its application in plasma physics (see, e.g., Ref.~\cite{diamond2010modern}). Indeed, in all of our models, both sudden and slow, the prediction $|P_1|=|P_0|$ is correctly verified, with the only exception of the asymmetric model SR. The reason is that for this model the evolution passes through a point where $P_0\simeq 0$. In this case, the real part of the eigenfrequency $\omega_0=-P_0\simeq 0$, so there is a breaking of adiabaticity where the eigenmodes grow faster than they oscillate. The WKB approximation cannot be applied, and therefore the predictions of quasi-linear theory cannot be trusted, and indeed they fail for this special case.

{\bf\textit{Discussion.}}---The present work was prompted by the large hierarchy of timescales between the mechanisms that drive a system towards an unstable state, and the much more rapid relaxation induced by FFC. Thus, starting from a strongly unstable configuration is never representative for a real physical system. Our numerical experiments have shown that, indeed, if the mechanism inducing the instability is slow, the asymptotic final state can differ significantly compared to the ``sudden'' appearance of the instability, a more conventional scenario in the literature. In both cases, the system tends to stick to the closest configuration which is stable in linear analysis; qualitatively, for our two-beam setup this corresponds to equipartition for one of the beams. However, the amplitudes of the oscillations around this state are much smaller for the more physical slow approach to the instability. Most importantly, the history of how the instability develops influences even the average properties of the final state if there is a crossing of a configuration with $P^z_0=0$, as in our SR model.

The idea of the instability as driven by a slowly-varying external agent was previously discussed in Refs.~\cite{Johns:2023jjt,Johns:2024dbe} in the context of miscidynamics, a thermodynamical theory of the neutrino final state. The main assumption of this theory is that the space-average of the neutrino density matrix would relax to a thermal form. Under this assumption, the external change was proposed to act as an adiabatic transformation on the neutrino flavor configuration. Our treatment is independent of this assumption, which at present remains speculative, and looks at the practical evolution of the system in the slowly varying setup. The separation of the neutrino flavor field into a spatially slowly-varying and a rapidly-varying part is common both to our quasi-linear approach and to the miscidynamics one, but the two approaches differ in the properties of the slowly-varying part, which we do not predict or assume to be thermalized. 

In numerical simulations, the role of a slow driving mechanism is sometimes implicit in the collisional term slowly forging the neutrino angular distribution, see e.g.\ Refs.~\cite{Capozzi:2018clo,Xiong:2024tac}. Here we reduce the driving mechanism to its most schematic features, to obtain a clean setup where its role comes out the clearest. Our conclusion is that the system always stays close to linear stability.

This simple conclusion is consistent with previous findings that the conversions tend to remove the angular crossing~\cite{Nagakura:2022xwe,Nagakura:2022kic,Zaizen:2022cik, Zaizen:2023ihz, Xiong:2023vcm, Cornelius:2023eop,Nagakura:2023jfi}, which in our language corresponds to one of the two beams reaching equipartition. However, at this stage this statement should be regarded as empirical; despite sounding natural, it is actually highly nontrivial, since the system tends towards a linearly stable state after entering nonlinear dynamics. For the first time, we provide an analytical explanation of this phenomenon, using quasi-linear theory typically applied to plasma systems~\cite{Vedenov1962quasi, drummond1961non, Drummond:1964}. 
For the cases where $P^z_0=0$, quasi-linear theory cannot be applied, because the unstable eigenmodes are not precessing at all, and indeed our numerical experiments show that for this case full equipartition is not obtained in either of the two beams. For the two-beam case, the outcome of this theory is nontrivial but very simple, since the stable angular distribution always corresponds to one of the two beams in equipartition. 

An extension of this theory for a continuous distribution of neutrino velocities would be of great practical use, describing the nonlinear evolution of the systems close to stability in an essentially analytical way. We leave this challenging question for future work, which could potentially shed light on what is the exact final state driven by fast conversions in these simple systems.

{\bf\textit{Acknowledgments.}}---We thank Marie Cornelius,  Basudeb Dasgupta, Lucas Johns, Hiroki Nagakura, Shashank Shalgar, G{\"u}nter Sigl, Irene Tamborra, and Zewei Xiong for discussions or comments on the manuscript. We thank the Galileo Galilei Institute for Theoretical Physics for the hospitality and the INFN for partial support during the revision of this work. DFGF is supported by the Villum Fonden under Project No.\ 29388 and the European Union's Horizon 2020 Research and Innovation Program under the Marie Sk{\l}odowska-Curie Grant Agreement No.\ 847523 ``INTERACTIONS.'' GGR acknowledges partial support by the German Research Foundation (DFG) through the Collaborative Research Centre ``Neutrinos and Dark Matter in Astro- and Particle Physics (NDM),'' Grant SFB-1258-283604770, and under Germany’s Excellence Strategy through the Cluster of Excellence ORIGINS EXC-2094-390783311.

\onecolumngrid
\appendix

\setcounter{equation}{0}
\setcounter{figure}{0}
\setcounter{table}{0}
\setcounter{page}{1}
\makeatletter
\renewcommand{\theequation}{S\arabic{equation}}
\renewcommand{\thefigure}{S\arabic{figure}}
\renewcommand{\thepage}{S\arabic{page}}

\begin{center}
\textbf{\large Supplemental Material for the Letter\\[0.5ex]
{\em Fast flavor conversions at the edge of instability in a two-beam model}}
\end{center}

We briefly introduce quasi-linear methods in plasma physics, where they were first applied, and we provide more details on the derivation of the quasi-linear equations of motion used in the main text.

\bigskip

\twocolumngrid

\section{A.~Quasi-linear saturation of plasma instabilities.}

In this section, we review the historical derivation of the quasi-linear treatment of plasma instabilities to provide some introductory background. The physical origin of this instability is the spontaneous emission of Cherenkov waves from resonant electrons moving in phase with the wave, as we have reviewed in Refs.~\cite{Fiorillo:2024bzm, Fiorillo:2024uki}. The system is thus described by the coupled equations of motion for the electron distribution function and the electric field self-consistently generated by the electrons. We write the phase-space distribution function for the electrons $F(t,\br,\bv)$; its evolution is ruled by the Vlasov equation
\begin{equation}
    \partial_t F+\bv\cdot\partial_\br F+\frac{e \bE[F]}{m_e}\cdot\partial_\bv F=0,
\end{equation}
where $e$ and $m_e$ are the electron charge and mass respectively, and $\bE[F]$ is the electric field associated with the distribution function, to be obtained from the Poisson equation
\begin{equation}
    \partial_{\br}\cdot \bE[F]= e \int d^3\bv\, F(r,\br,\bv)-e n_i,
\end{equation}
where the last term on the right is the charge density of the ions, exactly canceling the total charge of the electrons.

The quasi-linear scheme consists of
a separation in the distribution function $F(t,\br,\bv)=n_e\left[f_0(t,\bv)+f(t,\br,\bv)\right]$, similar to the linearized analysis we reviewed in Ref.~\cite{Fiorillo:2024bzm}. However, the physical nature of the separation is now different; the space-dependent part $f$ is not necessarily small -- although it needs to be if the quasi-linear approximation is to be accurate -- but is separated by spatial averaging, such that the average over small distances $\langle F \rangle=n_e f_0$. The separation operator using the spatial average depends on the assumption of homogeneity of the initial unperturbed condition, but may more generally be realized as an average over the ensemble of the initial perturbations seeding the instability; the results are identical in this case. Notice that the electric field is produced only from the fluctuation, i.e.\ $\bE[n f_0]=0$, since the charge of the homogeneous electron distribution is exactly canceled by the ion charge density.

After replacing this expansion in the Vlasov equation, we can now separate the terms slowly varying in space, and the ones rapidly varying. Among the former, we must include the spatial average of the non-linear term, so that we may write
\begin{equation}\label{eq:non_fluctuating}
    \partial_t f_0+\frac{e n_e}{m_e}\partial_\bv\cdot \langle \bE[f]f\rangle=0.
\end{equation}

In the rapidly-varying terms, we are going to keep only the terms linear in $f$, in agreement with the assumptions of quasi-linear theory. Therefore, we recover the linear equations also obtained in Ref.~\cite{Fiorillo:2024bzm}
\begin{equation}
    \partial_t f +\bv\cdot\partial_\br f+\frac{e n_e}{m_e}\partial_\bv f_0 \cdot \bE[f]=0.
\end{equation}
We expand this in spatial Fourier components as $f=\sum_\bk f_\bk e^{i\bk\cdot\br}$, and introduce the notation $\bE[f_\bk]=\bE_\bk$, so that
\begin{equation}\label{eq:fluctuation}
    \partial_t f_\bv+i\bk\cdot\bv f_\bk+\frac{e n_e}{m_e}\partial_\bv f_0\cdot\bE_\bk=0.
\end{equation}

In the quasi-linear framework, the function $f_0$ changes in time. However, we now assume that this temporal evolution is much slower than the characteristic frequencies of the fluctuating part $f$. Therefore, at every instant we can treat $f_0$ as approximately constant, and use the results reviewed in Ref.~\cite{Fiorillo:2024bzm}. The asymptotic temporal $f_\bk$ evolution is then determined by the characteristic frequencies $\omega_\bk$ that solve the dispersion relation
\begin{equation}
    1=-\frac{\omega_{\rm P}^2}{\bk^2}\int d^3\bv \frac{\bk\cdot\partial_\bv f_0(\bv)}{\omega_\bk-\bv\cdot\bk},
\end{equation}
where $\omega_{\rm P}$ is the plasma frequency and $\omega_\bk=\omega^0_\bk+i\gamma_\bk$ is generally complex. These frequencies determine the asymptotic temporal evolution. From Eq.~\eqref{eq:fluctuation}, we find
\begin{equation}
    f_\bk=-\frac{ien_e}{m_e(\omega_\bk-\bk\cdot\bv+i\epsilon)}\bE_\bk\cdot\partial_\bv f_0,
\end{equation}
where the $i\epsilon$ prescription is introduced explicitly to maintain causality. We now replace this solution in Eq.~\eqref{eq:non_fluctuating} and perform the spatial average. Assuming isotropy, we may use
\begin{equation}
    \langle E^i_\bk E^j_{\bk'}\rangle= \bE^2_\bk \delta_{ij} \delta_{\bk,\bk'},
\end{equation}
where $i$ is the spatial component of the vector $\bE_\bk$. Therefore, we find
\begin{equation}
    \partial_t f_0 -\partial_\bv\cdot\left[D(\bv) \partial_\bv f_0\right]=0,
\end{equation}
where we have introduced the diffusion coefficient
\begin{equation}
    D(\bv)=\frac{i e^2 n_e^2}{m_e^2}\sum_\bk \frac{\bE^2_\bk}{\omega_\bk-\bk\cdot\bv+i\epsilon}.
\end{equation}
In the sum, we can keep only the unstable eigenmodes, which are the ones that grow in time to a sizable degree; furthermore, since the result must be real, we can write
\begin{eqnarray}
    D(\bv)&=&\frac{D(\bv)+D^*(\bv)}{2}
    \nonumber\\
    &=&\frac{e^2 n_e^2}{m_e^2}\sum_\bk \bE_\bk^2 \frac{\gamma_\bk}{(\omega_\bk^0-\bk\cdot\bv)^2+\gamma_\bk^2}.
\end{eqnarray}
This equation must be complemented by a corresponding one for the time evolution of the fluctuating electric field
\begin{equation}
    \partial_t \bE_\bk^2=2\gamma_\bk \bE_\bk^2.
\end{equation}
Together, these equations provide a complete description of the saturation of the instability in the quasi-linear approximation. We do not review here the application of these equations to a concrete problem, e.g. the historical solution of the bump-on-tail instability~\cite{Vedenov1962quasi, drummond1961non, Drummond:1964}; we refer to textbook references, e.g.\ Ref.~\cite{schKT}, for a more detailed treatment of these topics.

\section{B.~Detailed derivation of the quasi-linear equations}

We now provide more details on the derivation of the quasi-linear equations discussed in the main text. We start from the exact equations of motion (EOMs) for the polarization vectors introduced in the main text and here reproduced for convenience
\begin{subequations}\label{eq:EOM}
    \begin{eqnarray}\label{eq:EOM-R}
    \partial_t \vP_{\rm R}+\partial_r \vP_{\rm R}&=&\bigl(\vP_0-\vP_1\bigr)\times \vP_{\rm R},
    \\ \label{eq:EOM-L}
    \partial_t \vP_{\rm L}-\partial_r \vP_{\rm L}&=&\bigl(\vP_0+\vP_1\bigr)\times \vP_{\rm L}.
    \end{eqnarray}
\end{subequations}
These equations can be spelled out in components, separating the longitudinal ones $P^z_{\rm L, R}$ and the transverse ones, the latter conveniently packed into a single complex variable $\psi_{\rm L, R}=P^x_{\rm L,R}+iP^y_{\rm L,R}$. As in the main text, we introduce the two moments of the distributions as $\vec{P}_{0,1}=\vec{P}_{\rm L}\pm\vec{P}_{\rm R}$, and analogously for the separate components $P^z$ and $\psi$, finding
\begin{subequations}\label{eq:fluct}
\begin{eqnarray}
    \partial_t \psi_0+\partial_r\psi_1&=&0,\\ 
    \partial_t \psi_1+\partial_r\psi_0&=&2i\left(P^z_0 \psi_1-P^z_1 \psi_0\right),
\end{eqnarray}    
\end{subequations}
and
\begin{subequations}
\begin{eqnarray}
    \kern-2em\partial_t P^z_0+\partial_r P^z_1&=&0,
    \\ 
    \kern-2em\partial_t P^z_1+\partial_r P^z_0&=&i\left(\psi_0 \psi_1^*-\psi_1 \psi_0^* \right).
\end{eqnarray}    
\end{subequations}
The Fourier components $\tilde{\vP}_{\rm L,R}=\int_0^L dr \vP_{\rm L,R}(t,r) e^{-ikr}$ then obey the equations
\begin{subequations}\label{eq:transv_component_eom}
\begin{eqnarray}
\kern-2em\partial_t\tpsi_0+ik\tpsi_1&=&0,\\[1ex]
    \kern-2em\partial_t \tpsi_1+ik\tpsi_0&=&2i\sum_{k'}\Bigl[P^z_0(t,k') \tpsi_1(t,k-k')
    \nonumber\\[-1ex]
    \kern-2em&&\kern2em{}-P^z_1(t,k') \tpsi_0(t,k-k')\Bigr],
\end{eqnarray}    
\end{subequations}
and 
\begin{subequations}\label{eq:z_component_eom}
\begin{eqnarray}
    \kern-2em\partial_t P^z_0+ik P^z_1&=&0,
    \\ [1ex]
    \kern-2em\partial_t P^z_1+ikP^z_0&=&i\sum_{k'}\Bigl[\tpsi_1^*(t,k') \tpsi_0(t,k'-k)
    \nonumber\\[-1ex]
    \kern-2em&&\kern2em{}-\tpsi_0^*(t,k') \tpsi_1(t,k'-k)\Bigr].
\end{eqnarray}    
\end{subequations}
These equations are exact. The crux of quasi-linear theory is the separation of the solution into a slowly-varying background and a fluctuating part. In our problem, given the symmetries of the initial condition, the background solution can be obtained directly as the spatial average of the $z$ component of the polarization vectors, so we introduce $P_{0,1}(t)=\langle P_{0,1}^z (t,r)\rangle$, which coincides with the Fourier component with $k=0$. The fluctuating part of the solution, in quasi-linear theory, is taken to be small enough that it can be linearized. Therefore, it coincides with the transverse components $\psi_{0,1}(t,r)$ only, since the fluctuating part of the $z$ components is of higher order.

After performing this separation, we can now write the EOM for $P_{0,1}$ as the $k=0$ component of Eqs.~\eqref{eq:z_component_eom}, obtaining the equations reported in the main text
\begin{subequations}
\begin{eqnarray}
    \kern-3em\partial_t P_0&=&0,
    \\ 
    \kern-3em\partial_t P_1&=&i\sum_{k}\left[\tpsi_0(t,k) \tpsi_1^*(t,k)-\tpsi_0^*(t,k) \tpsi_1(t,k)\right].
\end{eqnarray}    
\end{subequations}
For the fluctuating part, we only need to replace in Eq.~\eqref{eq:transv_component_eom} the $z$ components $P^z_{0,1}(t,r)$ with their background value $P_{0,1}$, obtaining
\begin{subequations}\label{eq:linearized_eom}
\begin{eqnarray}
    \partial_t\tpsi_0+ik\tpsi_1&=&0,\\ 
    \partial_t \tpsi_1+ik\tpsi_0&=&2i\left[P_0 \tpsi_1-P_1\tpsi_0\right].
\end{eqnarray}    
\end{subequations}

This finally leads to the quasi-linear EOM reported in the main text, where we detail our approximate method of solution following the same strategy traditionally applied in plasma physics and reviewed in Sect.~A.


\bibliographystyle{bibi.bst}
\bibliography{References}

\end{document}